\documentclass[aps,showpacs]{revtex4}
\usepackage{graphicx}
\def\bc{\begin{center}}
\def\ec{\end{center}}

\def\beq{\begin{equation}}
\def\eeq{\end{equation}}

\def\d{\downarrow}
\def\u{\uparrow}

\begin{document}

%Title of paper
\title{Short note on the Rabi model\\
%\small{(rabi solution.tex)}
}

\author{K. Ziegler}
\affiliation{Institut f\"ur Physik, Universit\"at Augsburg, D-86135 Augsburg, Germany\\
{\rm and}\\
Physics Department, New York City College of Technology, CUNY, %The City University of New York,
Brooklyn, NY 11201, USA}
\date{\today}

\begin{abstract}
The spectral density of the Rabi model is calculated exactly within a continued fraction approach.
It is shown that the method yields a convergent solution. 
\end{abstract}


\maketitle

A recent work by Braak \cite{braak11} has renewed the interest in the old problem of coupling 
a photon field to a single spin 1/2 state, using the Rabi model \cite{rabi37}. 
The central statement of this work is that the eigenfunctions in Bargmann representation must be
analytic functions in the entire complex plane. Based on this condition, a procedure is derived 
from the series expansion of the eigenstates which provides a recursive evaluation of the spectrum.
It is claimed that the series expansion yields an exact solution of the Rabi model,
which cannot be obtained from a direct continued fraction (CF) approach. In the following 
it is shown that this statement is not valid, and that the use of the extra condition of
analyticity of the eigenfunction in Bargmann representation is not necessary. Therefore,
the CF is directly applicable to the Rabi model and yields an exact solution for the spectral
density, where the term ``exact'' means that the CF is convergent and the spectral density can be evaluated 
by a simple algorithm to any desired accuracy.

The Rabi model is defined by the Hamiltonian
\beq
H_R=\omega a^\dagger a+\Delta\sigma_3 + U(a^\dagger +a)\sigma_1
\ ,
\label{hamiltonian00}
\eeq
where $a^\dagger$ ($a$) are creation (annihilation) operators of a photon,
the Pauli matrices $\sigma_j$ ($j=0,...,3$) describe operations on the spin-1/2 state,
and $\Delta$ is a symmetry breaking field for the spin. $U$ is the coupling strength between
the photons and the spin states.
This Hamiltonian maps a product state $|N\rangle\otimes |\sigma\rangle$ with $N$ photons
to $|N\pm 1\rangle\otimes |\sigma'\rangle$, where $\sigma,\sigma'=\d,\u$.
Thus eigenstates of $H$ are superpositions of product states, which can be derived
by a recursive approach for the coefficients of the superposition \cite{schweber67}. 
In the following we apply the recursive projection method (RPM) of Ref. \cite{ziegler11}.
This method is based on a random walk expansion in the underlying Hilbert
space which visits each subspace at most once \cite{ziegler10a}. 
Consequently, there are no loops in the random walk, which leads directly 
to a CF.

The RPM, as described in Ref. \cite{ziegler11}, can be directly applied to the 
resolvent $\langle N;s |(z-H)^{-1}|N;s'\rangle$ and yields immediately the
spectral density with respect to $|N;s\rangle$:
\beq
\rho_N(E)=\frac{1}{2}
\sum_{s=\uparrow,\downarrow}\lim_{\epsilon\to0}Im(\langle N;s |(E-i\epsilon-H)^{-1}|N;s\rangle)
\ .
\label{density}
\eeq
The simplest case is the resolvent $\langle 0;s |(z-H)^{-1}|0;s'\rangle$ 
of states without photons $|0,s\rangle$ and with spin projection $s=\uparrow,\downarrow$: 
\beq
(\langle 0;s |(z-H)^{-1}|0;s'\rangle)=\lim_{n\to \infty}\pmatrix{
g_n & 0 \cr
0 & h_n \cr
}
\ ,
\label{limit}
\eeq
where the eigenvalues of $H$ are the poles of the resolvent.
Then the matrix elements $g_n$, $h_n$ are subject to the following recurrence relations
\beq
g_k=\frac{1}{z-\omega(n-k)+\Delta-U^2(n-k+1) h_{k-1}}
\eeq
%, \ \ \
\beq
h_k=\frac{1}{z-\omega(n-k)-\Delta-U^2(n-k+1) g_{k-1}} % \ \ \ (1\le k\le n)
\eeq
for $1\le k\le n$ with the initial values
\beq
g_0=\frac{1}{z-\omega n-\Delta}\ , \ \ \ h_0=\frac{1}{z-\omega n+\Delta}
\ .
\eeq
The iteration of these recurrence relations gives a finite CF of the standard form \cite{perron12}
\beq
g_n=b_0+\frac{a_1}{b_1+\frac{a_2}{b_2+...}}
\equiv b_0+\frac{a_1|}{|b_1}+\cdots +\frac{a_n|}{|b_n}
\label{cf00}
\eeq
with coefficients
\beq
a_1=1, \ \ \ a_k=(1-k)U^2, \ \ \ b_0=0, \ \ \ b_1=z-\Delta,\ \ \ b_k=z-(k-1)\omega+(-1)^k\Delta
\ \ \ (2\le k\le n)
\ .
\label{coeffcients}
\eeq
For the corresponding coefficients of $h_n$ we must only replace $\Delta$ by $-\Delta$.
Therefore, all considerations for $g_n$ apply to $h_n$ after replacing $\Delta\to -\Delta$.
Due to $b_0=0$ the inverse of $g_n$ can be obtained directly by 
inverting $g_n$ in Eq. (\ref{cf00}):
\beq
a_1/g_n
%=b_1+\frac{a_2}{b_2+\frac{a_3}{b_3+...}}
=b_1+\frac{a_2|}{|b_2}+\cdots +\frac{a_n|}{|b_n}
\ .
\label{icf00}
\eeq
This means that the poles of $g_n$ are obtained from the zeros of the related CF,
and vice versa.

By taking the limit $g=\lim_{n\to\infty}g_n$ the CF in Eq. (\ref{cf00}) or (\ref{icf00})
yields the spectral density of the Rabi model for the complex energy $z$. This 
requires the limit $n\to\infty$ though, whose existence is a consequence of the Pringsheim 
Theorem (cf. \cite{perron12}): Considering the tail $t_n$ of the CF
\beq
g=b_0+\frac{a_1|}{|b_1}+\cdots +\frac{a_n|}{|b_n}+t_{n}
\eeq
with
\beq
t_{n}=
%\frac{a_{n+1}}{b_{n+1}+\frac{a_{n+2}}{b_{n+2}+...}}
\frac{a_{n+1}|}{|b_{n+1}}+\cdots
\ ,
\label{tail}
\eeq
we estimate the error when we truncate $g$ at $k=n$ (which gives $g_n$).
The Pringsheim Theorem states that for complex $a_k$, $b_k$ with
$|b_k|\ge |a_k|+1$ ($k\ge n$) the truncated CF $t_{n,l}$
\beq
t_{n,l}=\frac{a_{n+1}|}{|b_{n+1}}+\cdots +\frac{a_{l}|}{|b_{l}} \ \ \  (n+1< l)
\eeq
is convergent for $l\to\infty$ and converges to a $t_n$ with 
\beq
0\le |t_n|\le 1
\ .
\label{tail1}
\eeq
For the coefficients of the Rabi model in Eq. (\ref{coeffcients}) the conditions for the coefficients
are satisfied for a given $z$ and for sufficiently large $n$ if $U^2<\omega$. 
Then (\ref{tail1}) means that the ratio $t_n/b_n$ is $O(1/n)$, which implies
\beq
t_{n-1}=\frac{a_{n}}{b_{n}+t_{n}}=\frac{a_{n}}{b_{n}}+O(1/n)
=\frac{(1-n)U^2}{z-(n-1)\omega-(-1)^n\Delta}+O(1/n)
=\frac{U^2}{\omega}+O(1/n)
\ .
\label{tail2}
\eeq
This result enables us to approach the exact solution with a sequence of truncated CF's
with a given error $O(1/n)$:
\beq
g=b_0+\frac{a_1|}{|b_1}+\cdots +\frac{a_{n-1}|}{|b_{n-1}}+\frac{a_n/b_n+O(1/n)|}{|1}
\ .
\eeq
The finite CF can also be expressed as the ratio
\beq
g=\frac{A_{n-1}+[a_n/b_n+O(1/n)]A_{n-2}}{B_{n-1}+[a_n/b_n+O(1/n)]B_{n-2}}
\ ,
\label{final}
\eeq
where $A_{n-l}$ ($B_{n-l}$) are polynomials in $z$ of order $n-l-1$ ($n-l$) which are 
generated by the recurrence relations (cf. \cite{perron12})
\beq
A_k=b_kA_{k-1}+a_kA_{k-2}, \ \ \ B_k=b_kB_{k-1}+a_kB_{k-2} \ \ \ (k\ge1)
\eeq
with initial conditions $A_{-1}=1$, $A_0=b_0=0$, $B_{-1}=0$, $B_0=1$.

%\subsection{Conclusions}

{\it Conclusions and remarks:}
Eq. (\ref{final}) allows us to approach successively the spectral density
of the Rabi Hamiltonian, where the accuracy is improved with increasing $n$.
The error is estimated by the tail $t_{n}$ of (\ref{tail2}).
According to our definition this provides an exact solution of the Rabi spectrum,
where the poles of $g$ and $h$ are the corresponding eigenvalues.
The evaluation of $\rho_N(E)$ for $N>0$ can be performed within the same approach.  

The individual matrix elements $g$, $h$ avoid level crossing due to parity conservation, since
eigenstates of consecutive eigenvalues have different parity. However, the sum of $g_n$ and $h_n$ 
in the spectral density $\rho_0(E)$ provides level crossing by shifting their levels
relative to each other, for instance, by changing $\Delta$.

The ``rotating-wave'' approximation of the Rabi Hamiltonian yields the Jaynes-Cummings Hamiltonian
\beq
H_{JC}=\omega a^\dagger a+\Delta\sigma_3 + U(a^\dagger\sigma_- +a\sigma_+)
\eeq
with $\sigma_\pm=(\sigma_1\pm i\sigma_2)/2$. This Hamiltonian does not belong to the class of
models with recursive solutions because the recurrence relation of the RPM terminates
already after the first step for each $k$:
\[
g_k=\frac{z-\omega (n-k+1)-\Delta}
{[z-\omega (n-k+1)]^2-\Delta^2+\omega-U^2(n-k+1)}
\]
\beq
h_k=\frac{1}{z-\omega (n-k)-\Delta}
\ .
\eeq
Here the evaluation of the eigenvalues requires only the solution of a 
quadratic equation for each $k$, in agreement with the original work on
this model \cite{jaynes63,cummings65}.
\begin{figure}
\begin{center}
\includegraphics[width=10cm,height=7cm]{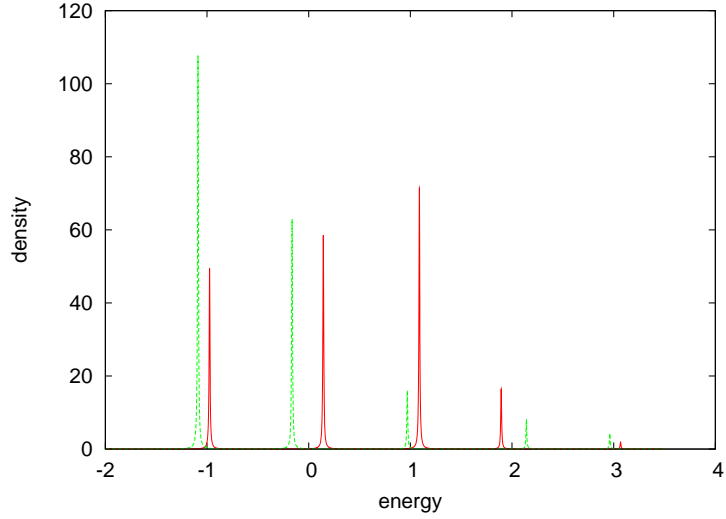}
\caption{
Spectral density $\rho_0(E)$ for maximally $n=500$ photons and 
%$\omega=1$, 
$\Delta=0.4$, $U=0.99$, $\epsilon=0.005$ in units of the photon frequency
$\omega$. The green (red) curves are contributions from $g_n$ ($h_n$).
}
\label{fig:1}
\end{center}
\end{figure}

\acknowledgments
The author is grateful for the hospitality at the New York City College of Technology 
during his sabbatical.

\end{document}